\documentclass[sigconf]{acmart}

\usepackage[english]{babel}
\usepackage[free-standing-units,per-mode=repeated-symbol,binary-units=true,detect-weight=true,detect-family=true]{siunitx}[=v2]
\usepackage{import}
\usepackage{glossaries}
\usepackage{cleveref}
\usepackage{enumitem}
\usepackage{eqparbox}
\usepackage{etoolbox}
\usepackage{pgfplots}
\usepackage{tikz}
\usepackage{pgf-pie}
\usepackage{subcaption}
\usepackage{enumitem}
\usepackage[referable]{threeparttablex}

\copyrightyear{2022}
\acmYear{2022}
\setcopyright{acmcopyright}
\acmConference[ICCAD '22]{IEEE/ACM International Conference on Computer-Aided Design}{October 30-November 3, 2022}{San Diego, CA, USA}
\acmBooktitle{IEEE/ACM International Conference on Computer-Aided Design (ICCAD '22), October 30-November 3, 2022, San Diego, CA, USA}
\acmPrice{15.00}
\acmDOI{10.1145/3508352.3549367}
\acmISBN{978-1-4503-9217-4/22/10}

\acmSubmissionID{356}

\acmConference[ICCAD'22]{International Conference on Computer-Aided Design}{30 October--3 November, 2022}{San Diego, CA, USA}
\acmYear{2022}

\newacronym[longplural={Scratchpad Memories}]{SPM}{SPM}{Scratchpad Memory}
\newacronym{ACE}{ACE}{AXI Coherent Extensions}
\newacronym{SCM}{SCM}{Standard Cell Memory}
\newacronym{LMUL}{LMUL}{Length Multiplier}
\newacronym{AMBA}{AMBA}{Advanced Microcontroller Bus Architecture}
\newacronym{VAU}{VAU}{Vector Arithmetic Unit}
\newacronym{APB}{APB}{Advanced Peripheral Bus}
\newacronym{API}{API}{Application Programming Interface}
\newacronym{ASIC}{ASIC}{Application-Specific Integrated Circuit}
\newacronym{AVX}{AVX}{Advanced Vector Extension}
\newacronym{AXI}{AXI}{Advanced eXtensible Interface}
\newacronym{BLAS}{BLAS}{Basic Linear Algebra Subprograms}
\newacronym{CHI}{CHI}{Coherent Hub Interface}
\newacronym{CMOS}{CMOS}{Complementary Metal-Oxide-Semiconductor}
\newacronym{CNN}{CNN}{Convolutional Neural Network}
\newacronym{CPU}{CPU}{Central Processing Unit}
\newacronym{CSR}{CSR}{Control and State Register}
\newacronym{CTS}{CTS}{Clock Tree Synthesis}
\newacronym{DLP}{DLP}{Data Level Parallelism}
\newacronym{DMA}{DMA}{Direct Memory Access}
\newacronym{DRAM}{DRAM}{Dynamic Random-Access Memory}
\newacronym{3R1W}{3R1W}{three read ports and one write port}
\newacronym{DSP}{DSP}{Digital Signal Processing}
\newacronym{DUT}{DUT}{Device Under Test}
\newacronym{ECL}{ECL}{Emitter-Coupled Logic}
\newacronym{FBB}{FBB}{Forward Body-Biasing}
\newacronym{FDSOI}{FD-SOI}{Fully Depleted Silicon on Insulator}
\newacronym{FMA}{FMA}{Fused Multiply-Add}
\newacronym{FPGA}{FPGA}{Field-Pro\-gram\-ma\-ble Gate Array}
\newacronym{FPU}{FPU}{Floating Point Unit}
\newacronym{GPGPU}{GPGPU}{General-Purpose \acrlong{GPU}}
\newacronym{GPU}{GPU}{Graphics Processing Unit}
\newacronym{HDL}{HDL}{Hardware Description Language}
\newacronym{HERO}{HERO}{Heterogeneous Embedded Research Platform}
\newacronym{HPC}{HPC}{High-Performance Computing}
\newacronym{IDol}{I\$}{Instruction Cache}
\newacronym{ILP}{ILP}{Instruction Level Parallelism}
\newacronym{IOT}{IoT}{Internet-of-Things}
\newacronym{IPC}{IPC}{Instructions Per Cycle}
\newacronym{IPU}{IPU}{Image Processing Unit}
\newacronym{MACU}{MACU}{Multiply-Accumulate Unit}
\newacronym{ISA}{ISA}{Instruction Set Architecture}
\newacronym{LSU}{LSU}{Load/Store Unit}
\newacronym{LVT}{LVT}{low voltage threshold}
\newacronym{MIMD}{MIMD}{Multiple Instruction, Multiple Data}
\newacronym{MMU}{MMU}{Memory Management Unit}
\newacronym{MUL}{MUL}{multiplier}
\newacronym{MVL}{MVL}{maximum vector length}
\newacronym{NUMA}{NUMA}{non-uniform memory access}
\newacronym{NOC}{NoC}{Network-on-Chip}
\newacronym{PCIe}{PCIe}{Peripheral Component Interconnect Express}
\newacronym{PC}{PC}{Program Counter}
\newacronym{PE}{PE}{Processing Element}
\newacronym{PL}{PL}{Programmable Logic}
\newacronym{PMCA}{PMCA}{Programmable Manycore Accelerator}
\newacronym{PSL}{PSL}{Power Service Layer}
\newacronym{PTE}{PTE}{page-table entry}
\newacronym{PTW}{PTW}{page-table walker}
\newacronym{PULP}{PULP}{Parallel Ultra Low Power}
\newacronym{RAW}{RAW}{read-after-write}
\newacronym{RBB}{RBB}{Reverse Body-Biasing}
\newacronym{ROB}{ROB}{Reorder Buffer}
\newacronym{RTL}{RTL}{Register Transfer Level}
\newacronym{RVT}{RVT}{Regular Voltage Threshold}
\newacronym{RoCC}{RoCC}{Rocket Custom Coprocessor Interface}
\newacronym{SIMD}{SIMD}{Single Instruction, Multiple Data}
\newacronym{SIMT}{SIMT}{Single Instruction, Multiple Thread}
\newacronym{SLDU}{SLDU}{Slide Unit}
\newacronym{SLVT}{SLVT}{super-low voltage threshold}
\newacronym{SM}{SM}{Streaming Multiprocessor}
\newacronym{CMG}{CMG}{Core Memory Group}
\newacronym{RAM}{RAM}{Random-Access Memory}
\newacronym{SRAM}{SRAM}{Static Random-Access Memory}
\newacronym{SSE}{SSE}{Streaming SIMD Extension}
\newacronym{SVE}{SVE}{Scalable Vector Extension}
\newacronym{MVE}{MVE}{M-Profile Vector Extension}
\newacronym{TLP}{TLP}{Thread Level Parallelism}
\newacronym{RVV}{RVV}{RISC-V Vector Extension}
\newacronym{TxnID}{TxnID}{Transaction ID}
\newacronym{VAC}{VAC}{Vector Access}
\newacronym{VC}{VC}{virtual channel}
\newacronym{VCONV}{VCONV}{Vector Conversion}
\newacronym{VEX}{VEX}{Vector Execute}
\newacronym{VFU}{VFU}{vector functional unit}
\newacronym{VID}{VID}{Vector Instruction Decode}
\newacronym{VIS}{VISSUE}{Vector Instruction Issue}
\newacronym{VLIW}{VLIW}{Very Long Instruction Word}
\newacronym{VLOOP}{VLOOP}{Vector Loop}
\newacronym{VLR}{VLR}{vector length register}
\newacronym{VLSU}{VLSU}{Vector Load/Store Unit}
\newacronym{VSLDU}{VSLDU}{Vector Slide Unit}
\newacronym{CC}{CC}{Core Complex}
\newacronym{VNB}{VNB}{Von Neumann Bottleneck}
\newacronym{VRF}{VRF}{Vector Register File}
\newacronym{VT}{VT}{vector thread}
\newacronym{WAR}{WAR}{write-after-read}
\newacronym{WAW}{WAW}{write-after-write}
\newacronym{DCT}{DCT}{discrete cosine transform}
\newacronym{TSV}{TSV}{through-silicon via}
\newacronym{3DIC}{3D-IC}{three-dimensional integrated circuit}
\newacronym{PPA}{PPA}{power, performance, and area}
\newacronym{F2F}{F2F}{face-to-face}
\newacronym{W2W}{W2W}{wafer-to-wafer}
\newacronym{IC}{IC}{integrated circuit}
\newacronym{C4}{C4}{controlled collapse chip connection}
\newacronym{FEOL}{FEOL}{front end of the line}
\newacronym{BEOL}{BEOL}{back end of the line}
\newacronym{PDP}{PDP}{power-delay product}
\newacronym{EDP}{EDP}{energy-delay product}
\newacronym{DRV}{DRV}{design rule violation}
\newacronym{DDR}{DDR}{double data rate}
\newacronym{SDRAM}{SDRAM}{synchronous dynamic random-access memory}

\definecolor{MidnightBlue}{HTML}{191970}
\definecolor{Mint}{HTML}{3EB889}
\definecolor{EnglishRed}{HTML}{A4515C}
\definecolor{SelectiveYellow}{HTML}{FFBA08}
\definecolor{CyanProcess}{HTML}{08B2E3}
\definecolor{OliveDrab7}{HTML}{4D4730}
\definecolor{Red}{HTML}{FF0000}

\colorlet{color1}{MidnightBlue}
\colorlet{color2}{Mint}
\colorlet{color3}{EnglishRed}
\colorlet{color4}{SelectiveYellow}
\colorlet{color5}{CyanProcess}
\colorlet{color6}{OliveDrab7}
\colorlet{colorAlert}{Red}

\DeclareSIUnit\flop{FLOP}
\DeclareSIUnit\flops{FLOPS}
\DeclareSIUnit\gate{GE}
\DeclareSIUnit\op{OP}
\DeclareSIUnit\macu{MACU}
\DeclareSIUnit\ops{OPS}
\DeclareSIUnit\cycle{cycle}
\DeclareSIUnit[number-unit-product = ]\percent{\%}

\newcommand\eg{e.g.,\ }
\newcommand\ie{i.e.,\ }
\newcommand\etal{et\penalty50\ al.\ }
\newcommand\mempoolspatz[2]{\ensuremath{\text{MemPool}_{#1}\text{Spatz}_{#2}}}
\newcommand\mempool[1]{\ensuremath{\text{MemPool}_{#1}}}
\newcommand\spatz[1]{\ensuremath{\text{Spatz}_{#1}}}
\newcommand\bigO[1]{\ensuremath{\mathcal{O}(#1)}}

\renewlist{tablenotes}{enumerate}{1}
\makeatletter
\setlist[tablenotes]{label=\tnote{(\alph*)},ref={(\alph*)},itemsep=\z@,topsep=\z@skip,partopsep=\z@skip,parsep=\z@,itemindent=\z@,labelindent=\tabcolsep,labelsep=.2em,leftmargin=*,align=left,before={\footnotesize}}
\makeatother

\newlist{rdescription}{description}{1}
\AtBeginEnvironment{rdescription}{%
\renewcommand*\descriptionlabel[2][Des]{\hspace\labelsep\eqmakebox[Des][r]{\hfill\normalfont\bfseries #2}}\setlist[rdescription]{leftmargin=\dimexpr\eqboxwidth{Des}+\labelsep}}%

\usetikzlibrary{scopes, calc, shapes, arrows, positioning, patterns}
\tikzset{>=latex}

\pgfplotsset{compat=1.13}
\pgfplotsset{width=\linewidth, height=7cm}
\pgfplotsset{every x tick label/.append style={font=\small}}
\pgfplotsset{every y tick label/.append style={font=\small}}

\pgfplotsset{
  /pgf/declare function={
    roof(\x,\b,\p) = (\b * \x < \p) * \b * \x + (\b * \x >= \p) * \p;}}

\usepgfplotslibrary{external}
\usepgfplotslibrary{groupplots}
\usepgfplotslibrary{fillbetween}

\begin{document}


\title{Spatz: A Compact Vector Processing Unit for High-Performance and Energy-Efficient Shared-L1 Clusters}

\author{Matheus Cavalcante}
\authornote{Both authors contributed equally to this research.}
\email{matheus@iis.ee.ethz.ch}
\author{Domenic W\"uthrich}
\authornotemark[1]
\email{domenicw@student.ee.ethz.ch}
\affiliation{%
  \institution{Integrated Systems Laboratory\\ETH Z\"urich}
  \city{Z\"urich}
  \country{Switzerland}
}

\author{Matteo Perotti}
\email{mperotti@iis.ee.ethz.ch}
\author{Samuel Riedel}
\email{sriedel@iis.ee.ethz.ch}
\affiliation{%
  \institution{Integrated Systems Laboratory\\ETH Z\"urich}
  \city{Z\"urich}
  \country{Switzerland}
}

\author{Luca Benini}
\email{lbenini@iis.ee.ethz.ch}
\affiliation{%
  \institution{Integrated Systems Laboratory\\ETH Z\"urich}
  \city{Z\"urich}
  \country{Switzerland}
}
\affiliation{%
  \institution{Universit\`a di Bologna}
  \city{Bologna}
  \country{Italy}
}

\renewcommand{\shortauthors}{Cavalcante and W\"uthrich, et al.}

\begin{abstract}
  While parallel architectures based on clusters of \glspl{PE} sharing
  L1 memory are widespread, there is no consensus on how lean their
  \gls{PE} should be. Architecting \glspl{PE} as vector processors
  holds the promise to greatly reduce their instruction fetch
  bandwidth, mitigating the \gls{VNB}. However, due to their
  historical association with supercomputers, classical vector
  machines include microarchitectural tricks to improve the \gls{ILP},
  which increases their instruction fetch and decode energy
  overhead. In this paper, we explore for the first time vector
  processing as an option to build small and efficient \glspl{PE} for
  large-scale shared-L1 clusters. We propose Spatz, a compact, modular
  32-bit vector processing unit based on the integer embedded subset
  of the RISC-V Vector Extension version 1.0. A Spatz-based cluster
  with four \glspl{MACU} needs only \SI{7.9}{\pico\joule} per $32$-bit
  integer multiply-accumulate operation, \SI{40}{\percent} less energy
  than an equivalent cluster built with four Snitch scalar cores. We
  analyzed Spatz' performance by integrating it within MemPool, a
  large-scale many-core shared-L1 cluster. The Spatz-based MemPool
  system achieves up to \SI{285}{\giga\ops} when running a
  \num{256x256} $32$-bit integer matrix multiplication,
  \SI{70}{\percent} more than the equivalent Snitch-based MemPool
  system. In terms of energy efficiency, the Spatz-based MemPool
  system achieves up to \SI{266}{\giga\ops\per\watt} when running the
  same kernel, more than twice the energy efficiency of the
  Snitch-based MemPool system, which reaches
  \SI{128}{\giga\ops\per\watt}. Those results show the viability of
  lean vector processors as high-performance and energy-efficient
  \glspl{PE} for large-scale clusters with tightly-coupled L1 memory.
\end{abstract}



\maketitle


\glsresetall{}

\section{Introduction}
\label{sec:introduction}

The ever-growing need for computing performance under an increasingly
limited power budget is the defining characteristic of modern computer
architecture. As an answer to the phase-out of Moore's
Law~\cite{Moore1975} and Dennard's scaling~\cite{Dennard1974},
computer architects must strive for improved scalability and energy
efficiency to propel performance scaling in the post-Moore
era~\cite{Vetter2018}. This challenge has led to an architectural
shift from exploiting high \gls{ILP} towards the exploitation of
on-chip \gls{MIMD} parallelism~\cite{Domke2022}.

A common architectural pattern comprises clusters of \glspl{PE} that
share tightly-coupled L1 memory through a low-latency
interconnect~\cite{MemPool2021}. \glspl{GPU}, which dominate most of
the Top500 list~\cite{Top5002021}, follow this architectural
pattern. For example, NVIDIA Hopper \glspl{GPU} are composed of
several \glspl{SM} with four tensor cores and \SI{256}{\kibi\byte} of
shared L1 data cache. Each tensor core controls several \gls{FMA}
units, with the whole \gls{SM} capable of $1024$ FP16/FP32 \gls{FMA}
operations per cycle~\cite{NvidiaH1002020}.

Despite the diffusion of shared-L1 \gls{PE} clusters, there is no
consensus on how small their \gls{PE} should be. However, evidence
shows that ultra-small cores controlling large functional units can be
energetically
efficient~\cite{Zaruba2020}. Manticore~\cite{Zaruba2020b} took this
approach, using the tiny (\SI{22}{\kilo\gate}) Snitch
core~\cite{Zaruba2020} to build a $4096$-core system composed of four
compute chiplets. Each chiplet contains $128$ clusters, each featuring
\SI{128}{\kibi\byte} of tightly-coupled L1 memory and eight small
Snitch cores equipped with large double-precision \glspl{FPU}. Another
example is MemPool~\cite{MemPool2021}, a scaled-up Snitch-based Snitch
with $256$ cores sharing \SI{1}{\mebi\byte} of L1 \gls{SPM} accessible
by all cores within at most five cycles of zero-load latency. However,
there is a fundamental trade-off between the number of \glspl{PE} and
the \gls{VNB}, \ie the memory traffic and energy overhead due to the
instruction fetching mechanism and related
logic~\cite{Backus1978}. Therefore, maximizing the \gls{PE}'s
instruction fetch efficiency is the key challenge for improving the
overall system's energy efficiency.

Instead of scaling by exploiting \gls{MIMD}, the \gls{SIMD}
parallelism tackles the \gls{VNB} by executing the same instruction on
several chunks of data. In particular, the vector-\gls{SIMD} approach
promises to reach very high performance and energy efficiency numbers
without requiring the ultra-wide datapaths of packed-\gls{SIMD}-based
architectures~\cite{Kozyrakis2003b}. Moreover, by exploiting
\gls{DLP}, vector engines are potentially the most efficient approach
to tackle the \gls{VNB}.

From its inception with the Cray-1 machine to the modern
A64FX~\cite{Yoshida2018}, vector processing has always been associated
with supercomputers. Vector processors usually include all
microarchitectural tricks to increase \gls{ILP}, \eg renaming,
out-of-order execution, speculation, and branch prediction, which
increase the area and energy overhead of classical high-performance
vector processors. However, the defining characteristic of a vector
processor is not \gls{ILP} but \gls{DLP}. This key observation has led
to the design of more streamlined vector cores where most hardware
resources are dedicated to \gls{DLP} support, \ie a wide \gls{VRF}
with parallel execution lanes~\cite{Ara2020}. In this vein, the idea
of an embedded vector machine is gaining traction with modern vector
\glspl{ISA}. Arm's \gls{MVE}~\cite{Armv81M2019} and the \verb#Zve*#
subset of the \gls{RVV}~\cite{RISCV2022} both target small vector
machines for edge data-parallel processing, as opposed to
high-performance computing.

Ultimately, the key benefit of a vector \gls{ISA} is that the fetch
and decode cost of a single vector instruction can be amortized over
many data-processing cycles. In this paper, we explore for the first
time vector processing as an option to build small and efficient
\glspl{PE} for large-scale clusters with tightly-coupled L1 memory. We
propose \emph{Spatz}, a compact $32$-bit vector machine based on the
embedded subset of the RISC-V Vector Extension version
1.0~\cite{RISCV2022}. We use Spatz as a building block to improve the
performance and efficiency of a many-core cluster. The contributions
of this paper are:
\begin{itemize}
\item The physically-driven microarchitectural design of Spatz, a
  parametric, small 32-bit vector unit based on the \verb#Zve32x#
  subset of \gls{RVV} version 1.0. Spatz uses a generic accelerator
  interface, allowing it to work in tandem with any scalar core
  compatible with this interface (\Cref{sec:architecture});
\item A performance analysis of Spatz and Spatz-based MemPool
  instances on key data-parallel kernels. The resulting system is
  highly scalable, achieving the roofline boundary of reachable
  performance on a wide range of kernels and core-count configurations
  (\Cref{sec:spatz-based-shared,sec:performance});
\item The architectural exploration and physical implementation of
  Spatz with post-place-and-route results in the modern
  GlobalFoundries' 22FDX \gls{FDSOI} technology. We compare the
  baseline MemPool design with the Spatz-based MemPools in terms of
  area, power consumption, energy efficiency, and area efficiency. The
  Spatz-powered MemPool reaches \SI{285}{\giga\ops}, \SI{71}{\percent}
  higher than the baseline design, all with a power consumption
  \SI{13}{\percent} lower. By using Spatz as its \gls{PE}, we more
  than double MemPool's energy efficiency, reaching
  \SI{266}{\giga\ops\per\watt} (\Cref{sec:phys-impl});
\item Insights about using a small vector-processor-based \gls{PE} as
  the building block of a large shared-L1 many-core system in terms of
  scalability, performance, energy efficiency, and programmability
  (\Cref{sec:related-work,sec:conclusion}).
\end{itemize}

Our design is open-sourced under a liberal license\footnote{Footnote
  hidden for double-blind review purposes.}.


\section{Architecture}
\label{sec:architecture}

Spatz is a small parametric vector unit based on the \acrfull{RVV}
version 1.0, supporting instructions from the \gls{RVV} \verb#Zve32x#
subset for embedded vector machines. This section describes Spatz'
architecture, highlighting its main components.

\begin{figure}[htb]
  \centering
  \includegraphics[width=.9\linewidth]{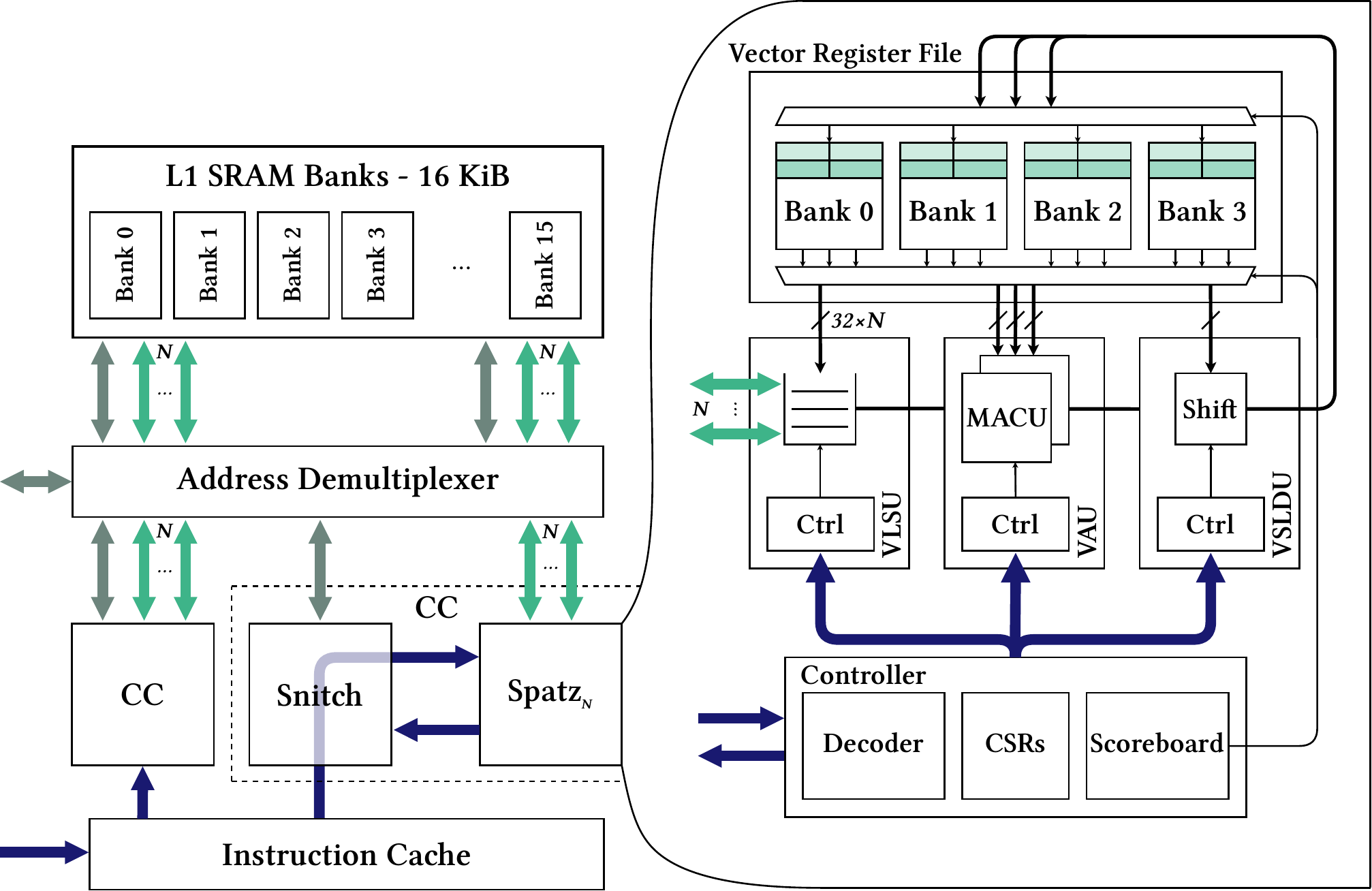}
  \caption{Microarchitecture of \spatz{N}, a Spatz instance with $N$
    \glspl{MACU}, and its integration in a small shared-L1 cluster.}
  \label{fig:spatz_arch}
\end{figure}

\Cref{fig:spatz_arch} details the microarchitecture of \spatz{N}, a
Spatz instance with $N$ \glspl{MACU}, and its integration within the
MemPool tile. Spatz has a latch-based \gls{VRF}, divided into four
banks with \gls{3R1W}. The \gls{VRF} serves data to Spatz' functional
units. Spatz' controller is responsible for keeping track of the
in-flight vector instructions and the interface with the scalar core,
in our case Snitch, which is responsible for executing scalar
instructions and forwarding vector instructions to Spatz. Snitch and
Spatz form a \gls{CC}.

\spatz{N} is integrated within a small-scale shared-L1 cluster with
\SI{16}{\kibi\byte} of local \gls{SPM}, divided into $16$ \gls{SRAM}
banks with \SI{1}{\kibi\byte} each. Each \gls{CC} has a private
latch-based L0 \gls{IDol} of \SI{128}{\byte} and share
\SI{2}{\kibi\byte} of L1 \gls{IDol}. An address-based demultiplexer
decides whether the \gls{CC} memory requests are forwarded to the
\gls{AXI} interface or the logarithmic crossbar between the \glspl{CC}
and L1 \gls{SPM} banks.

\subsection{Instruction Dispatch}
\label{sec:instruction-dispatch}

Spatz implements a subset of the \gls{RVV} \gls{ISA}, version
1.0~\cite{RISCV2022}. Notably, we target the \verb#Zve32x# subset,
designed for small embedded vector machines with 8-bit, 16-bit, and
32-bit integer support. Out of the \verb#Zve32x# subset, Spatz does
not currently support vector reductions or scatter-gather
operations. However, our centralized \gls{VRF} means that adding
support for vector permutation instructions should not affect Spatz'
microarchitecture (\Cref{sec:vector-register-file}). Further extending
Spatz towards full compliance with \verb#Zve32x# is on our roadmap.

Spatz is processor-agnostic. It communicates with the scalar core
through the generic CORE-V X-Interface accelerator
interface~\cite{OpenHW2022}. Therefore, Spatz can interface with any
core compatible with the X-Interface. Our choice of using Snitch is
justified by its extremely lightweight footprint, adapted for a vector
execution paradigm where most of the computation happens in the vector
machine.

Unfortunately, since the interface specification is still in its
infancy, it was not well-adapted for the memory bandwidth requirements
of a vector machine. We extended the X-Interface to consider cases
where the accelerator makes its memory accesses through a memory
interface much wider than the scalar cores' one. We guarantee ordering
between Spatz and Snitch memory requests by stalling the scalar core's
Load/Store Unit while Spatz' Vector Load/Store Unit executes a memory
operation and vice-versa.

\subsection{Controller}
\label{sec:controller}

Snitch only pre-decodes vector instructions, dispatching the vector
instruction and any scalar operands to the vector unit. Spatz'
controller decodes the vector instructions, keeps track of their
execution, and acknowledges their completion with Snitch.

The controller also manages the \glspl{CSR} of the \gls{RVV}
\gls{ISA}. For example, the \verb#vlen# \gls{CSR} defines the vector
length of all vector instructions. Another important \gls{CSR} is
\verb#vtype#, which controls the vector elements' width and whether
physical vector registers are grouped into longer logical vector
registers. This grouping, called \gls{LMUL}, allows for a logical
vector length up to eight times longer than the machines' vector
length, at the expense of fewer available logic vector registers.

Finally, the controller orchestrates the execution of the vector
instructions in the functional units. The scoreboard keeps track of
how many elements of each vector instruction were committed into the
\gls{VRF}. Hazards between vector instructions are handled through
operand backpressure. Spatz supports chaining by calculating the
hazards on a per-element basis.

\subsection{Vector Register File}
\label{sec:vector-register-file}

The \gls{VRF} is the heart of any vector machine. In Spatz, we decided
on a multi-banked multi-ported \gls{VRF} with four \gls{3R1W}
banks. Each bank is implemented as a latch-based \gls{SCM}. Spatz'
\gls{VRF} is also centralized and serves all Spatz' functional
units. The \gls{VRF} ports match the throughput requirements of the
\verb#vmacc# instruction, which reads three operands to produce one
result. Each bank is $\verb#VLEN#/4$ bits wide, and each of the $32$
\verb#VLEN#-bit-wide vector registers occupies one row in the four
\gls{VRF} banks. Each \gls{VRF} port is $32N$ bits wide, where $N$ is
the number of \glspl{MACU} in Spatz.

Despite the scaling issues typically associated with multi-ported
\glspl{VRF}~\cite{Ara2020}, the \gls{VRF} is not a limiting factor for
the target \gls{MACU} count of Spatz. Our vector unit is designed as
the \gls{PE} of a shared-L1 cluster, with a handful of \glspl{MACU}
per Spatz. We can further scale our design by replicating the
shared-L1 cluster, connecting them with a low-latency L1
interconnect~\cite{MemPool2021}. This replication has two
benefits. First, the resulting many-core system retains its \gls{MIMD}
flexibility. At the same time, the vector units tackle the
high-throughput and computing-intensive
tasks~\cite{Shintani2022}. Second, only very large problems can
efficiently exploit vector machines with large \gls{MACU}
count~\cite{Ara2020}; hence our \gls{MIMD} cluster achieves high
\gls{MACU} utilization even when the workload is not suitable for
super-wide vectors.

The centralized \gls{VRF} also helps the implementation of vector
permutation instructions, \eg a vector slide
($\verb#vd#[i] \leftarrow \verb#vs#[i \pm \verb#shamt#]$). A
lane-based vector architecture, where the vectors of a \gls{VRF} are
divided into lanes based on their index $i$, would need to shuffle the
elements and store them in the correct lane, with important
scalability implications~\cite{Ara2020}. Thanks to its centralized
\gls{VRF}, Spatz can efficiently implement vector slides with a barrel
shifter.

\subsection{Functional Units}
\label{sec:functional-units}

Spatz has three functional units: the \gls{VLSU}, the \gls{VAU}, and
the \gls{VSLDU}.

\subsubsection{\acrlong{VLSU}}
\label{sec:acrlongvlsu}

The \gls{VLSU} handles the memory interfaces of Spatz, with support
for unit-strided and constant-strided memory accesses. The \gls{VLSU}
supports a parametric number of $32$-bit wide memory interfaces. By
default, the number of memory interfaces $N$ matches the number of
\glspl{MACU} in the design. This implies a peak operation per memory
bandwidth ratio of \SI{0.5}{\op\per\byte}.

Spatz' independent and narrow memory interfaces allow the reuse of the
same $32$-bit wide L1 \gls{SPM} interconnect used by the scalar
cores. The independent interfaces also allow fast
constant-strided---and potentially scatter-gather---execution, as the
\gls{VLSU} does not need to coalesce requests into wide memory
transfers. However, since there is no ordering guarantee between the
memory responses of the individual requests, a \gls{ROB} sits between
the memory interfaces and the \gls{VRF}. The \gls{ROB} ensures that
the memory responses are written as ordered $32N$-bit-wide words to
the \gls{VRF}, simplifying the chaining mechanism in the scoreboard.

\subsubsection{\acrlong{VAU}}
\label{sec:acrlongvau}

\begin{figure}[ht]
  \centering
  \includegraphics[width=0.6\linewidth]{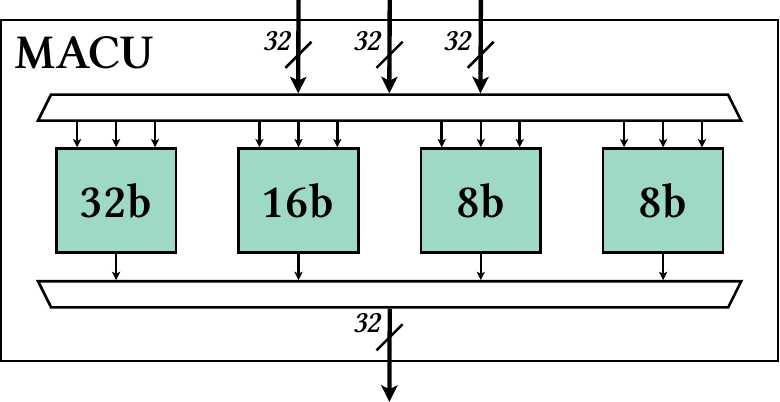}
  \caption{Architecture of Spatz' \gls{MACU}.}
  \label{fig:arch_vau}
\end{figure}

The \gls{VAU} is Spatz' main functional unit, hosting $N$
\glspl{MACU}. The \gls{MACU} supports $8$-bit, $16$-bit, and $32$-bit
elements. Each \gls{MACU} has a throughput of $32$ bits per cycle,
regardless of the element width. Within one \gls{MACU}, execution
happens in a packed-\gls{SIMD} fashion. \Cref{fig:arch_vau} shows the
architecture of one of those \glspl{MACU}. For area saving purposes,
the \gls{MACU} has four datapaths, one $32$-bit wide, one $16$-bit
wide, and two $8$-bit wide. Narrow operations reuse the wide
datapaths.

Each datapath implements a multiplier, an adder, comparators, and
shifters. The most complex operation supported by the \gls{MACU}'s
datapaths is \verb#vmacc#, a multiply-accumulate operation that
requires three source operands per result.

\subsubsection{\acrlong{VSLDU}}
\label{sec:acrlongvsldu}

The \gls{VSLDU} executes vector permutation instructions. Examples of
such instructions include vector slide up/down and vector moves. The
unit operates on two private $32N$-bit wide register banks. Between
those two register banks, an all-to-all interconnect allows the
implementation of any permutation. Common operations, \eg slides,
produce results at a $32N$ bits per cycle ratio, which matches Spatz'
other functional units' peak throughput. The register banks also play
a role similar to the \gls{ROB} of Spatz' \gls{VLSU}. Not only the
\gls{VSLDU} does double-buffering on those registers, but it also
ensures that the unit commits to the \gls{VRF} in $32N$-bit-wide
words, which simplifies the chaining calculation in the scoreboard by
increasing its granularity.


\section{Spatz-based processing element}
\label{sec:spatz-based-shared}

In this section, we analyze the performance of a Spatz-based \gls{PE}
with key data-parallel kernels. We consider two differently-sized
Spatz configurations, \spatz{2} and \spatz{4}. Their design parameters
are summarized in \Cref{tab:spatz}. All Spatz configurations were
designed for a peak operation per memory bandwidth ratio of
\SI{0.5}{\op\per\byte}.

\begin{table}[ht]
  \centering
  \caption{Spatz configurations.}
  \begin{tabular}[ht]{rll}
    \toprule
                                            & \textbf{\spatz{2}} & \textbf{\spatz{4}} \\\midrule
    \#\acrshortpl{MACU}                     & 2                  & 4                  \\
    Vector length [\si{\bit}]               & 256                & 512                \\
    \acrshort{VRF} size [\si{\kibi\byte}]   & 1                  & 2                  \\
    Peak performance [\si{\op\per\cycle}]   & 4                  & 8                  \\
    Memory bandwidth [\si{\byte\per\cycle}] & 8                  & 16                 \\\bottomrule
  \end{tabular}
  \label{tab:spatz}
\end{table}

We used the shared-L1 cluster of \Cref{fig:spatz_arch} as the smallest
cluster with which we can analyze Spatz' \gls{PPA}. All considered
clusters have four \glspl{MACU} in total, either in a single \spatz{4}
\gls{CC}, two \spatz{2} \glspl{CC}, or four scalar Snitch \glspl{CC}.

\subsection{Performance}
\label{sec:performance-1}

We benchmark a single Spatz unit with key compute-bound signal
processing kernels. Those kernels operate on matrices stored in local
low-latency L1 memory. A \gls{DMA} engine can be used to copy data
from higher memory levels into the L1 memory, while Spatz operates on
local data.

The \emph{matmul} kernel, the multiplication of two $n \times n$
matrices, is the prime example of a compute-bound kernel for large
matrices. In fact, its arithmetic intensity is \bigO{n}. Spatz'
\gls{VRF} allows us to tile the matrix multiplication in blocks much
larger than the blocks that would fit in the register file of a scalar
core. For example, using \gls{RVV}'s \glspl{LMUL}, we can group up to
eight logical vector registers into a single one. In \spatz{4}, this
is a physical vector of $4096$~bits, enough to fit $128$ $32$-bit
matrix elements. We also benchmarked our system with \emph{conv2d},
the 2D integer convolution kernel. This kernel also has a large amount
of data reuse, with its arithmetic intensity a function of the kernel
size $f \times f$. On a multi-core environment, each scalar core has
its copy of the convolution kernel $K$ to avoid unnecessary banking
conflicts in the L1 memory. Spatz implements the \emph{conv2d}
algorithm with optimized vector slides.

The performance results were extracted with a cycle-accurate \gls{RTL}
simulation of the target kernels. The roofline plot of
\Cref{fig:spatz_roofline} shows \spatz{2}'s and \spatz{4}'s
performance on the \emph{matmul} and \emph{conv2d} kernels, together
with their maximum achievable performance. \spatz{2} reaches an
almost-ideal \gls{MACU} utilization for the considered benchmarks. The
peak performance for the \emph{matmul} benchmark is
\SI{3.84}{\op\per\cycle} (\SI{96.0}{\percent}), and the peak
performance for the \emph{conv2d} benchmark is
\SI{3.95}{\op\per\cycle} (\SI{98.8}{\percent}). Large kernels also
reach very high performance on \spatz{4}. Its peak performance on the
\emph{matmul} kernel is \SI{7.67}{\op\per\cycle} (\SI{95.8}{\percent})
and \SI{7.78}{\op\per\cycle} (\SI{97.2}{\percent}) on the
\emph{conv2d} kernel.

\begin{figure}[ht]
  \centering
  \begin{tikzpicture}[every mark/.append style={mark size=3pt, very thick}]
    \begin{axis}[
      height              = 5.2cm,
      xlabel              = {Arithmetic intensity [\si{\op\per\byte}]},
      x label style       = {at={(0.5,-0.3)}},
      xmode               = log,
      log basis x         = 4,
      xmin                = .25,
      xmax                = 16,
      xticklabel style    = {rotate=90},
      extra x ticks       = {0.5, 1.333, 2.666, 5.333, 2.250, 12.250},
      extra x tick labels = {0.5, \emph{matmul}$_{8}$, \emph{matmul}$_{16}$, \emph{matmul}$_{32}$, \emph{conv2d}$_3$, \emph{conv2d}$_7$},
      ylabel              = {Performance [\si{\op\per\cycle}]},
      ymode               = log,
      log basis y         = 2,
      ymin                = 1.8,
      ymax                = 8.5,
      extra y ticks       = {3, 5, 6, 7},
      grid                = major,
      legend style        = {at={(0.95,0.05)}, anchor=south east, font=\small},
      log ticks with fixed point]

      \addlegendimage{only marks, thick, color2, mark=o}
      \addlegendimage{only marks, thick, color1, mark=+}
      \addlegendentry{\spatz{4}}
      \addlegendentry{\spatz{2}}

      \addplot[color2, very thick, domain=0.25:1, samples=201, name path=roofSpatz4M]{roof(x,16,8)};
      \addplot[color2, very thick, domain=1:64, name path=roofSpatz4C] {8};

      \addplot[color1, very thick, domain=0.25:1, samples=201, name path=roofSpatz2M]{roof(x,8,4)};
      \addplot[color1, very thick, domain=1:64, name path=roofSpatz2C] {4};

      \path[name path=xAxisM] (axis cs:0.25,1) -- (axis cs:1.0,1);
      \path[name path=xAxisC] (axis cs:1.0,1) -- (axis cs:32.0,1);
      \addplot[fill=color1, fill opacity=.05] fill between [of=xAxisM and roofSpatz2M];
      \addplot[fill=color1, fill opacity=.05] fill between [of=xAxisC and roofSpatz2C];
      \addplot[fill=color2, fill opacity=.05] fill between [of=roofSpatz2M and roofSpatz4M];
      \addplot[fill=color2, fill opacity=.05] fill between [of=roofSpatz2C and roofSpatz4C];

      \pgfplotstableread{results/spatz/spatz2}\loadedtable;
      \addplot [only marks, thick, mark=+, color1] table [x=Intensity, y=Performance] {\loadedtable};
      \pgfplotstableread{results/spatz/spatz4}\loadedtable;
      \addplot [only marks, thick, mark=o, color2] table [x=Intensity, y=Performance] {\loadedtable};
    \end{axis}
  \end{tikzpicture}
  \caption{Roofline plot for \spatz{2} and \spatz{4} \glspl{PE}
    running the \emph{matmul} and \emph{conv2d} benchmarks. The
    subscript numbers beside the benchmark names indicate the matrix
    sizes $n$ for \emph{matmul} and the kernel size $f$ for
    \emph{conv2d}.}
  \label{fig:spatz_roofline}
\end{figure}

\Cref{fig:spatz_roofline} also shows how gracefully the performance of
the vector unit scales for small problems. On \spatz{2}, even a tiny
\emph{matmul}$_8$ reaches \SI{3.54}{\op\per\cycle}
(\SI{88.6}{\percent}). In contrast, on \spatz{4}, we remark a
performance degradation for this kernel size, reaching
\SI{5.17}{\op\per\cycle}. This \gls{MACU} utilization of
\SI{64}{\percent} is a consequence of the short execution time of the
\verb#vmacc# instructions, which take two cycles to process a matrix
row of eight elements. As a result, Snitch must issue a new
\verb#vmacc# every two cycles to keep Spatz' pipelines full. However,
bookkeeping scalar instructions limit the \verb#vmacc# issue rate to
once every three cycles. This limitation translates into a diagonal
maximum performance boundary on the roofline plot~\cite{Ara2020}.

The roofline shows that a Spatz-based \gls{PE} can reach high
performance and functional unit utilization even for very small
kernels. Moreover, Spatz achieves this without needing a superscalar
or out-of-order core. The vector abstraction allows a
pseudo-double-issue behavior, with the \gls{VAU}, \gls{VLSU}, and
\gls{VSLDU} working in parallel, without the scalar core issuing more
than one instruction per cycle. \Cref{sec:performance} will analyze
how Spatz performs in a many-core system, with several instances
competing for L1 memory bandwidth.

\subsection{Synthesis results}
\label{sec:phys-impl-1}

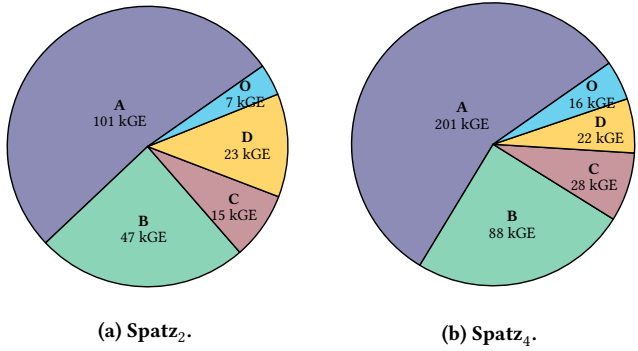
\begin{figure}[ht]
  \centering
  \begin{minipage}[h]{0.46\linewidth}
    \resizebox{\linewidth}{!}{
      \begin{tikzpicture}
        \pie [color = {color1!50, color2!60, color3!60, color4!60, color5!60}, sum=193, after number={~kGE}, radius=2.8, rotate=35, text=inside] {
          101/\textbf{A},
          47/\textbf{B},
          15/\textbf{C},
          23/\textbf{D},
          7/\textbf{O}
        }
      \end{tikzpicture}}
    \subcaption{\spatz{2}.}
    \label{fig:area_spatz2}
  \end{minipage}\hfill%
  \begin{minipage}[h]{0.46\linewidth}
    \resizebox{\linewidth}{!}{
      \begin{tikzpicture}
        \pie [color = {color1!50, color2!60, color3!60, color4!60, color5!60}, sum=355, after number={~kGE}, radius=2.8, rotate=35, text=inside] {
          201/\textbf{A},
          88/\textbf{B},
          28/\textbf{C},
          22/\textbf{D},
          16/\textbf{O}
        }
      \end{tikzpicture}}
    \subcaption{\spatz{4}.}
    \label{fig:area_spatz4}
  \end{minipage}
  \caption{Post-synthesis area distribution of \spatz{2} and \spatz{4}
    \glspl{CC}. The labels correspond to (A) \gls{VRF}; (B) \gls{VAU};
    (C) \gls{VLSU}; (D) Snitch; (O) other smaller blocks, \eg
    \gls{VSLDU} and controller.}
  \label{fig:spatz_area}
\end{figure}

We used Synopsys' Fusion Compiler 2022.03 to synthesize the Spatz- and
Snitch-based small-sized shared-L1 clusters of \Cref{fig:spatz_arch}
using GlobalFoundries' 22FDX \gls{FDSOI} technology. We target
\SI{500}{\mega\hertz} in worst-case conditions
(SS/\SI{0.72}{\volt}/\SI{125}{\celsius}). \Cref{fig:spatz_area} shows
the post-synthesis area distribution of the \spatz{2} and \spatz{4}
\glspl{CC}.

The \spatz{2}-based \gls{CC} is \SI{193}{\kilo\gate} large, or
\SI{97}{\kilo\gate\per\macu}. Snitch occupies \SI{12}{\percent} of it,
\SI{23}{\kilo\gate}, while \spatz{2} occupies the remaining
\SI{170}{\kilo\gate}. \spatz{2}'s \SI{1}{\kibi\byte}-large latch-based
\gls{VRF} occupies most of its footprint, \SI{101}{\kilo\gate}. The
\glspl{MACU} are the next largest component of \spatz{2}, occupying a
total of \SI{47}{\kilo\gate}. All remaining components have a very
small footprint. Notably, Spatz' controller and associated scoreboard
logic occupy only \SI{6}{\kilo\gate}. The \spatz{4}-based \gls{CC} is
\SI{355}{\kilo\gate} large, or \SI{89}{\kilo\gate\per\macu}. This
normalized footprint is \SI{8}{\percent} smaller than
\spatz{2}'s. Most of \spatz{4}'s footprint is occupied by the
\SI{2}{\kibi\byte}-large \gls{VRF}, which uses
\SI{201}{\kilo\gate}. This footprint is twice the footprint of
\spatz{2}'s \gls{VRF}, evidence of the scalability of our
architecture. \spatz{4} amortizes the footprint overhead of Spatz'
controller and Snitch, which contributes to the lower normalized
footprint of this \gls{CC} when compared to the \spatz{2}-based
\gls{CC}.

\begin{figure}[ht]
  \centering
  \begin{tikzpicture}[/tikz/font=\footnotesize]
    \begin{axis}[
      xbar stacked,
      xmax=11.5,
      bar width = 1em,
      y axis line style = {draw=none},
      axis x line = none,
      tickwidth = 0pt,
      height = 3.75cm,
      nodes near coords,
      legend style={at={(1.15,1)}, anchor=north east, cells={anchor=west}},
      symbolic y coords = {macc, mul, add, load},
      ytick distance = 1,
      yticklabel style={align=right, xshift=2ex, font=\footnotesize},
      every pin/.style={font=\footnotesize},
      nodes near coords align={center},
      point meta = explicit symbolic]

      \addplot+[xbar, color=black, fill=color4!60] plot coordinates {(0.10,macc) (0.10,mul) (0.10,add) (0.12,load)};
      \addplot+[xbar, color=black, fill=color1!50] plot coordinates {(2.89,macc) [\SI{2.9}{\pico\joule}] (2.16,mul) [\SI{2.2}{\pico\joule}] (2.21,add) [\SI{2.2}{\pico\joule}] (1.23,load) [\SI{1.2}{\pico\joule}]};
      \addplot+[xbar, color=black, fill=color2!60] plot coordinates {(2.77,macc) [\SI{2.8}{\pico\joule}] (2.53,mul) [\SI{2.5}{\pico\joule}] (2.59,add) [\SI{2.6}{\pico\joule}] (0.19,load)};
      \addplot+[xbar, color=black, fill=color3!60] plot coordinates {(0.05,macc) (0.05,mul) (0.05,add) (0.68,load)};
      \addplot+[xbar, color=black, fill=color5!60] plot coordinates {(0.92,macc) (0.93,mul) (0.91,add) (1.84,load) [\SI{1.8}{\pico\joule}]};
      \addplot+[xbar, color=black, fill=color6!50] plot coordinates {(1.103,macc) [\SI{1.1}{\pico\joule}] (1.103,mul) [\SI{1.1}{\pico\joule}] (1.103,add) [\SI{1.1}{\pico\joule}] (2.06,load) [\SI{2.1}{\pico\joule}]};

      \addplot+[color=black, nodes near coords align={right}] plot coordinates {(0.0001,macc) [\SI{7.9}{\pico\joule}] (0.0001,mul) [\SI{6.9}{\pico\joule}] (0.0001,add) [\SI{7.0}{\pico\joule}] (0.0001,load) [\SI{6.1}{\pico\joule}]};

      \legend{Snitch, \gls{VRF}, \gls{VAU}, \gls{VLSU}, Interconnect, \glspl{SRAM} (\gls{SPM}/\gls{IDol})}
    \end{axis}
  \end{tikzpicture}
  \caption{Breakdown of the \spatz{4}-based cluster's energy
    consumption per elementary operation in the vector instruction.}
  \label{fig:energy}
\end{figure}
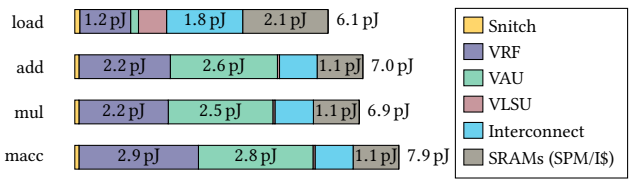

We used Synopsys' PrimePower 2022.03 to estimate the post-synthesis
energy consumption per elementary operation of the \spatz{4}-based
cluster in typical conditions (TT/\SI{0.80}{\volt}/\SI{25}{\celsius}),
using switching activities extracted from a gate-level
simulation. \Cref{fig:energy} shows this energy breakdown when running
\verb#vload#, \verb#vadd#, \verb#vmul#, or \verb#vmacc# instructions,
with a vector length of $32$ elements of $32$ bits.

Snitch only consumes \SI{0.12}{\pico\joule} per elementary operation
while issuing instructions to the vector unit. The low energy
requirement is because Snitch only needs to issue an instruction every
four cycles to keep \spatz{4}'s four \glspl{MACU} fully utilized. Also
of note is the energy efficiency gained from merging the
vector-multiply \verb#vmul# and vector-add \verb#vadd# instructions
into the vector-multiply-accumulate \verb#vmacc#
instruction. Successive \verb#vmul# and \verb#vadd# instructions
require \SI{13.9}{\pico\joule} per elementary multiply-accumulate
operation, whereas \verb#vmacc# requires \SI{36}{\percent} less energy
or \SI{7.9}{\pico\joule} on \spatz{4}. Spatz operates on data stored
in its \gls{VRF}, responsible for most of \spatz{4}'s energy
consumption. For example, \verb#vmacc# reads three operands and writes
one result in the \gls{VRF} per vector operation; the \gls{VRF}
requires \SI{2.9}{\pico\joule}, \SI{36}{\percent} of the cluster's
total energy consumption. The \gls{VAU} is the second major energy
consumer, with \SI{2.8}{\pico\joule} or \SI{35}{\percent} of the
overall consumption. The remaining energy is consumed by several
smaller blocks, \gls{SRAM} banks and interconnect.

\begin{figure}[ht]
  \centering
  \begin{tikzpicture}[/tikz/font=\footnotesize]
    \begin{axis}[
      xbar,
      xmin=0,
      xmax=14.5,
      enlargelimits=0.15,
      reverse legend,
      bar width = 0.7em,
      y axis line style = {draw=none},
      axis x line = none,
      tickwidth = 0pt,
      height = 5.5cm,
      nodes near coords,
      legend style={at={(1.05,1)}, anchor=north east, cells={anchor=west}},
      symbolic y coords = {macc, mul, add, load},
      ytick distance = 1,
      yticklabel style={align=right, xshift=4ex, font=\footnotesize},
      every pin/.style={font=\footnotesize},
      point meta=explicit symbolic,
      legend image code/.code={\draw [#1] (0cm,-0.1cm) rectangle (0.3cm,0.15cm);},
      nodes near coords style={
        anchor=west,
      },
      nodes near coords,
      nodes near coords align={horizontal}]

      \addplot+[xbar, color=black, fill=color4!60, postaction={pattern=north east lines}] plot coordinates {(13.1,macc) [\SI{13.1}{\pico\joule}] (12.9,mul) [\SI{12.9}{\pico\joule}] (8.0,add) [\SI{8.0}{\pico\joule}] (7.9,load) [\SI{7.9}{\pico\joule}]};

      \addplot+[xbar, color=black, fill=color2!60, postaction={pattern=north west lines}] plot coordinates {(7.9,macc) [\SI{7.9}{\pico\joule}] (6.9,mul) [\SI{6.9}{\pico\joule}] (7.0,add) [\SI{7.0}{\pico\joule}] (6.1,load) [\SI{6.1}{\pico\joule}]};

      \addplot+[xbar, color=black, fill=color1!50, postaction={pattern=crosshatch dots}] plot coordinates {(7.8,macc) [\SI{7.8}{\pico\joule}] (6.8,mul) [\SI{6.8}{\pico\joule}] (6.9,add) [\SI{6.9}{\pico\joule}] (6.6,load) [\SI{6.6}{\pico\joule}]};

      \legend{Snitch-based cluster, \spatz{4}-based cluster, \spatz{2}-based cluster}
    \end{axis}
  \end{tikzpicture}
  \caption{Energy consumption per elementary operation of
    \spatz{2}-based, \spatz{4}-based, and Snitch-based clusters
    running common instructions. All clusters instantiate 4
    \glspl{MACU}.}
  \label{fig:energy_comparison}
\end{figure}
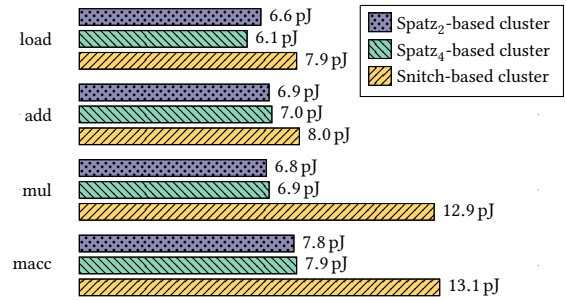

Both \spatz{2}-based and \spatz{4}-based clusters are more efficient
than the scalar Snitch-based cluster. \Cref{fig:energy_comparison}
compares the energy consumption per elementary operation of those
clusters. For example, the \spatz{4}-based cluster consumes
\SI{5.2}{\pico\joule} less energy than the Snitch-based cluster with
four scalar cores to run a multiply-accumulate instruction. This
\SI{40}{\percent} reduction in energy consumption highlights the
feasibility of embedded vector engines to mitigate the \gls{VNB} in
small shared-L1 clusters. Even for very simple arithmetic
instructions, such as an addition, the Spatz-based cluster requires
\SI{14}{\percent} less energy than the equivalent Snitch-based
cluster.

The \spatz{2}-based and \spatz{4}-based MemPool tiles have similar
energy consumption for the considered instructions. However, the
\spatz{2}-based tile has a footprint \SI{8}{\percent} larger, which
might impact its integration and replication at a higher hierarchy
level.


\section{Spatz-based MemPool Cluster}
\label{sec:performance}

We used Spatz as the \gls{PE} used to build up the MemPool many-core
system. In this section, we analyze this system's performance, in
\si{\op\per\cycle}, on key data-parallel kernels.

\subsection{MemPool configurations}
\label{sec:memp-conf}

MemPool~\cite{MemPool2021} is a highly-parametric design. Its smallest
unit, the tile (\Cref{fig:spatz_arch}), contains four Snitch cores,
\SI{2}{\kibi\byte} of L1 \gls{IDol}, \SI{16}{\kibi\byte} of \gls{SPM}
divided into $16$ \gls{SRAM} banks, and a fully-connected logarithmic
crossbar between the cores and memories. This tile can be replicated to
build systems with as low as $16$ cores (the ``\emph{minpool}''
configuration) to as high as $256$ cores (the ``\emph{mempool}''
configuration)~\cite{MemPool2021,MemPoolGitHub2021}. We analyze Spatz'
impact on the \gls{PPA} of several MemPool configurations,
representing a wide range of shared-L1 cluster sizes.

We name a specific MemPool configuration as \mempoolspatz{c}{i}, where
$c$ is the number of Snitch + Spatz cores in the system, and $i$ is
the number of \glspl{MACU} that each Spatz controls. A MemPool
configuration with scalar cores only is called \mempool{c}, where $c$
is the number of Snitches in the system, each controlling a single
\gls{MACU}. First, we define a few small \emph{minpool}
configurations, all with a peak performance of \SI{32}{\op\per\cycle}:
\begin{rdescription}
\item[\mempool{16}] Configuration with $16$ Snitch cores;
\item[\mempoolspatz{8}{2}] Configuration with $8$ Snitch + \spatz{2} cores;
\item[\mempoolspatz{4}{4}] Configuration with $4$ Snitch + \spatz{4} cores.
\end{rdescription}
Each Spatz instance has a \gls{VRF} with a vector length of
\SI{128}{\bit\per\macu}. In total, all \glspl{VRF} amount to
\SI{8}{\kibi\byte} of L0 storage.

We also define larger \emph{mempool} configurations with the
Snitch-only version corresponding to the largest MemPool instance
reported by Cavalcante~\etal\cite{MemPool2021}. All larger instances
have a peak performance of \SI{512}{\op\per\cycle}:
\begin{rdescription}
\item[\mempool{256}] Configuration with $256$ Snitch cores;
\item[\mempoolspatz{128}{2}] Configuration with $128$ Snitch + \spatz{2} cores;
\item[\mempoolspatz{64}{4}] Configuration with $64$ Snitch + \spatz{4} cores.
\end{rdescription}
Each Spatz instance has a \gls{VRF} with a vector length of
\SI{128}{\bit\per\macu}. In total, all \glspl{VRF} amount to
\SI{128}{\kibi\byte} of L0 storage.

\subsection{Benchmarks}
\label{sec:benchmarks}

\begin{figure}[ht]
  \centering
  \begin{tikzpicture}[every mark/.append style={mark size=3pt, very thick}]
    \begin{axis}[
      xlabel              = {Arithmetic intensity [\si{\op\per\byte}]},
      x label style       = {at={(0.5,-0.24)}},
      xmode               = log,
      log basis x         = 2,
      xmin                = 2,
      xmax                = 64,
      xticklabel style    = {rotate=90},
      xtick               = \empty,
      extra x ticks       = {2.666, 5.333, 10.666, 21.333, 42.666, 2.250},
      extra x tick labels = {\emph{matmul}$_{16}$, \emph{matmul}$_{32}$, \emph{matmul}$_{64}$, \emph{matmul}$_{128}$, \emph{matmul}$_{256}$, \emph{conv2d}$_3$},
      ylabel              = {Performance [\si{\op\per\cycle}]},
      ymode               = log,
      log basis y         = 2,
      ymin                = 8,
      ymax                = 600,
      extra y ticks       = {3},
      grid                = major,
      legend style        = {at={(1,-0.5)}, anchor=south east, legend columns=3, font=\small, /tikz/every even column/.append style={column sep=0.4cm}},
      log ticks with fixed point]

      \addlegendimage{only marks, thick, color2, mark=o}
      \addlegendimage{only marks, thick, color2, mark=+}
      \addlegendimage{only marks, thick, color2, mark=x}
      \addlegendimage{only marks, thick, color1, mark=o}
      \addlegendimage{only marks, thick, color1, mark=+}
      \addlegendimage{only marks, thick, color1, mark=x}

      \addlegendentry{\mempool{256}}
      \addlegendentry{\mempoolspatz{128}{2}}
      \addlegendentry{\mempoolspatz{64}{4}}
      \addlegendentry{\mempool{16}}
      \addlegendentry{\mempoolspatz{8}{2}}
      \addlegendentry{\mempoolspatz{4}{4}}

      \addplot[color2, very thick, domain=0.25:1, samples=201, name path=roofSpatz256M]{roof(x,512,512)};
      \addplot[color2, very thick, domain=1:64, name path=roofSpatz256C] {512};

      \addplot[color1, very thick, domain=0.25:1, samples=201, name path=roofSpatz16M]{roof(x,32,32)};
      \addplot[color1, very thick, domain=1:64, name path=roofSpatz16C] {32};

      \path[name path=xAxisM] (axis cs:0.5,1) -- (axis cs:1.0,1);
      \path[name path=xAxisC] (axis cs:1.0,1) -- (axis cs:64.0,1);
      \addplot[fill=color1, fill opacity=.05] fill between [of=xAxisM and roofSpatz16M];
      \addplot[fill=color1, fill opacity=.05] fill between [of=xAxisC and roofSpatz16C];
      \addplot[fill=color2, fill opacity=.05] fill between [of=roofSpatz16M and roofSpatz256M];
      \addplot[fill=color2, fill opacity=.05] fill between [of=roofSpatz16C and roofSpatz256C];

      \pgfplotstableread{results/mempool/vanilla}\loadedtable;
      \addplot [only marks, thick, mark=o, color2] table [x=Intensity, y=Performance] {\loadedtable};
      \pgfplotstableread{results/mempool/spatz2}\loadedtable;
      \addplot [only marks, thick, mark=+, color2] table [x=Intensity, y=Performance] {\loadedtable};
      \pgfplotstableread{results/mempool/spatz4}\loadedtable;
      \addplot [only marks, thick, mark=x, color2] table [x=Intensity, y=Performance] {\loadedtable};
      \pgfplotstableread{results/minpool/vanilla}\loadedtable;
      \addplot [only marks, thick, mark=o, color1] table [x=Intensity, y=Performance] {\loadedtable};
      \pgfplotstableread{results/minpool/spatz2}\loadedtable;
      \addplot [only marks, thick, mark=+, color1] table [x=Intensity, y=Performance] {\loadedtable};
      \pgfplotstableread{results/minpool/spatz4}\loadedtable;
      \addplot [only marks, thick, mark=x, color1] table [x=Intensity, y=Performance] {\loadedtable};
    \end{axis}
  \end{tikzpicture}
  \caption{Roofline plot for MemPool instances running the
    \emph{matmul} and \emph{conv2d} benchmarks. The subscript numbers
    besides the benchmark names indicate the matrix size $n$ for
    \emph{matmul} and the kernel size $f$ for \emph{conv2d}.}
  \label{fig:mempool_roofline}
\end{figure}
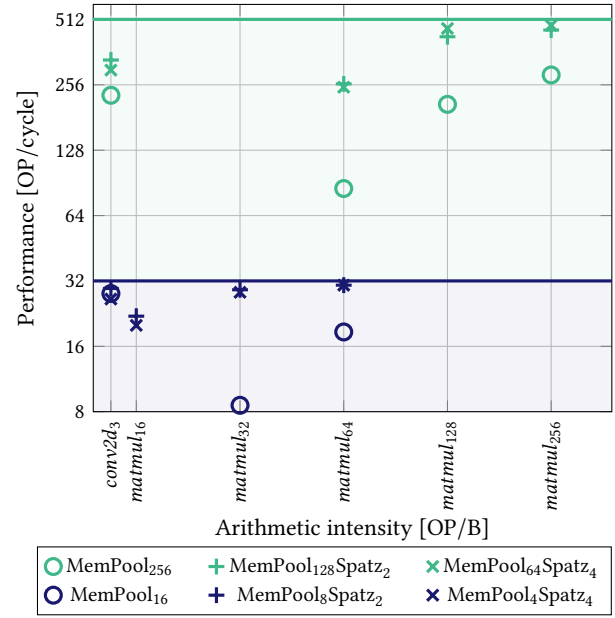

We benchmark the MemPool instances with the \emph{matmul} and
\emph{conv2d} compute-bound signal processing kernels. Our analysis is
performed assuming matrices stored in MemPool's low-latency L1 memory,
which is \SI{64}{\kibi\byte} large for the small \emph{minpool}
configurations, and \SI{1}{\mebi\byte} for the larger \emph{mempool}
configurations. \Cref{fig:mempool_roofline} shows the rooflines of the
considered MemPool configurations. The Spatz-based MemPool systems
reach equal or higher performance than the Snitch-based MemPool
systems for both the \emph{mempool} and \emph{minpool} configurations
and with all benchmarks.

Spatz' high performance and efficiency is fully confirmed in the small
MemPool configurations. The \mempoolspatz{8}{2} instance reaches
\SI{30.6}{\op\per\cycle} (\SI{95.6}{\percent}) on
\emph{matmul}$_{64}$. This performance is much higher than
\mempool{16}'s, \SI{18.6}{\op\per\cycle} (\SI{58.1}{\percent}). Spatz'
performance degrades slightly for small matrices. \mempoolspatz{8}{2}
achieves \SI{22.0}{\op\per\cycle} (\SI{68.8}{\percent}) on the
\emph{matmul}$_{16}$ kernel, with the performance being limited by
Snitch's issue rate. Despite reaching similar performance for large
problems, \mempoolspatz{4}{4} performs slightly worse than
\mempoolspatz{8}{2} on \emph{matmul}$_{16}$, reaching
\SI{20.0}{\op\per\cycle} (\SI{62.4}{\percent}) due to a different
\emph{matmul} tiling. The convolution also performs well on Spatz,
with \mempoolspatz{8}{2} reaching \SI{29.5}{\op\per\cycle}
(\SI{92.2}{\percent}) on \emph{conv2d}$_3$. \mempool{16}'s performance
is competitive with Spatz' performance, thanks to a handwritten
implementation of the \emph{conv2d}$_3$ kernel.

The advantage of a vector \gls{PE} grows with large MemPool
configurations. On \mempool{256}, for example, the
\emph{matmul}$_{256}$ kernel reaches \SI{284}{\op\per\cycle}
(\SI{55.4}{\percent}). Performance is limited by the data reuse and
the \emph{matmul} tiling, which is bounded by Snitch's scalar register
file. The \gls{VRF} allows for increased data reuse on Spatz, which
significantly improves Spatz' performance. Moreover, vector chaining
allows both the \gls{VLSU} and the \gls{VAU} to work concurrently, in
a pseudo-double-issue behavior. Thanks to that, \mempoolspatz{64}{4}
reaches \SI{480.2}{\op\per\cycle} (\SI{93.1}{\percent}) when running
\emph{matmul}$_{256}$. While there is some performance degradation for
smaller matrices---the \mempoolspatz{128}{2} instance reaches
\SI{258.3}{\op\per\cycle} (\SI{50.4}{\percent}) on
\emph{matmul}$_{64}$---this is still much higher than \mempool{256}'s
performance on the same kernel, which reaches \SI{85.2}{\op\per\cycle}
(\SI{16.6}{\percent}). With the convolution kernel,
\mempoolspatz{128}{2} reaches \SI{332.8}{\op\per\cycle}
(\SI{65.0}{\percent}) on \emph{conv2d}$_3$, higher than
\mempool{256}'s \SI{229}{\op\per\cycle} (\SI{44.7}{\percent}).


\section{Physical Implementation}
\label{sec:phys-impl}

This section analyzes the post-place-and-route \gls{PPA} metrics of
MemPool group instances that use either Snitch or Spatz as their
\glspl{PE}. We used Synopsys' Fusion Compiler 2022.03 to synthesize,
place, and route the MemPool groups using GlobalFoundries' advanced
22FDX \gls{FDSOI} technology. All designs target an operating
frequency of \SI{500}{\mega\hertz} in worst-case conditions
(SS/\SI{0.72}{\volt}/\SI{125}{\celsius}).

\subsection{Area and Performance}
\label{sec:area-performance}

\Cref{tab:mempool_group} summarizes the area and performance results
of MemPool groups derived from the \mempool{256},
\mempoolspatz{128}{2}, and \mempoolspatz{64}{4} configurations. On the
MemPool system, a \emph{group} contains a fourth of the cores and
\glspl{MACU}~\cite{MemPool2021}. Four identical groups form the
complete MemPool system, which only contains point-to-point
connections between four groups. The group is MemPool's most
timing-critical design, where most of the power is
consumed~\cite{MemPool2021}. In the following, we focus on the group
alone.

\begin{table*}[htbp]
  \centering
  \caption{Post-place-and-route \gls{PPA} results of \mempool{256}, \mempoolspatz{128}{2}, and \mempoolspatz{64}{4}.}
  \setlength{\tabcolsep}{12pt}
  \begin{threeparttable}
    \begin{tabular}[h]{rlll}
      \toprule
                                                                                    & \textbf{\mempool{256}} & \textbf{\mempoolspatz{128}{2}} & \textbf{\mempoolspatz{64}{4}} \\\midrule
      Area                         [\si{\milli\meter\squared}]                      & \num{15.8}             & \num{21.0}                     & \num{20.1}                    \\
      \emph{Area (MemPool group)} [\si{\milli\meter\squared}]                       & \num{2.75}             & \num{3.65}                     & \num{3.50}                    \\
      \emph{Cell Area Utilization (MemPool group)} [\si{\percent}]                  & \SI{68}{\percent}      & \SI{67}{\percent}              & \SI{66}{\percent}             \\\midrule
      Operating Frequency \emph{(worst-case)} [\si{\mega\hertz}]                    & \num{485}              & \num{471}                      & \num{472}                     \\
      Operating Frequency \emph{(typical)} [\si{\mega\hertz}]                       & \num{587}              & \num{591}                      & \num{594}                     \\\midrule
      Peak performance [\si{\giga\ops}]\tnotex{tnote:kernel}                        & \num{167}              & \num{270}                      & \num{285}                     \\
      Area efficiency [\si{\giga\ops\per\milli\meter\squared}]\tnotex{tnote:kernel} & \num{10.6}             & \num{12.0}                     & \num{14.2}                    \\\midrule
      Power consumption [\si{\watt}]\tnotex{tnote:kernel}                           & \num{1.30}             & \num{1.15}                     & \num{1.07}                    \\
      Energy efficiency [\si{\giga\ops\per\watt}]\tnotex{tnote:kernel}              & \num{128}              & \num{234}                      & \num{266}                     \\\bottomrule
    \end{tabular}
    \begin{tablenotes}
      \item\label{tnote:kernel} Extracted running the \emph{matmul}$_{256}$ kernel at typical operating conditions.
    \end{tablenotes}
  \end{threeparttable}
  \label{tab:mempool_group}
\end{table*}

\begin{figure}[ht]
  \centering
  \begin{minipage}[t]{0.5\linewidth}
    \centering
    \includegraphics[width=\linewidth]{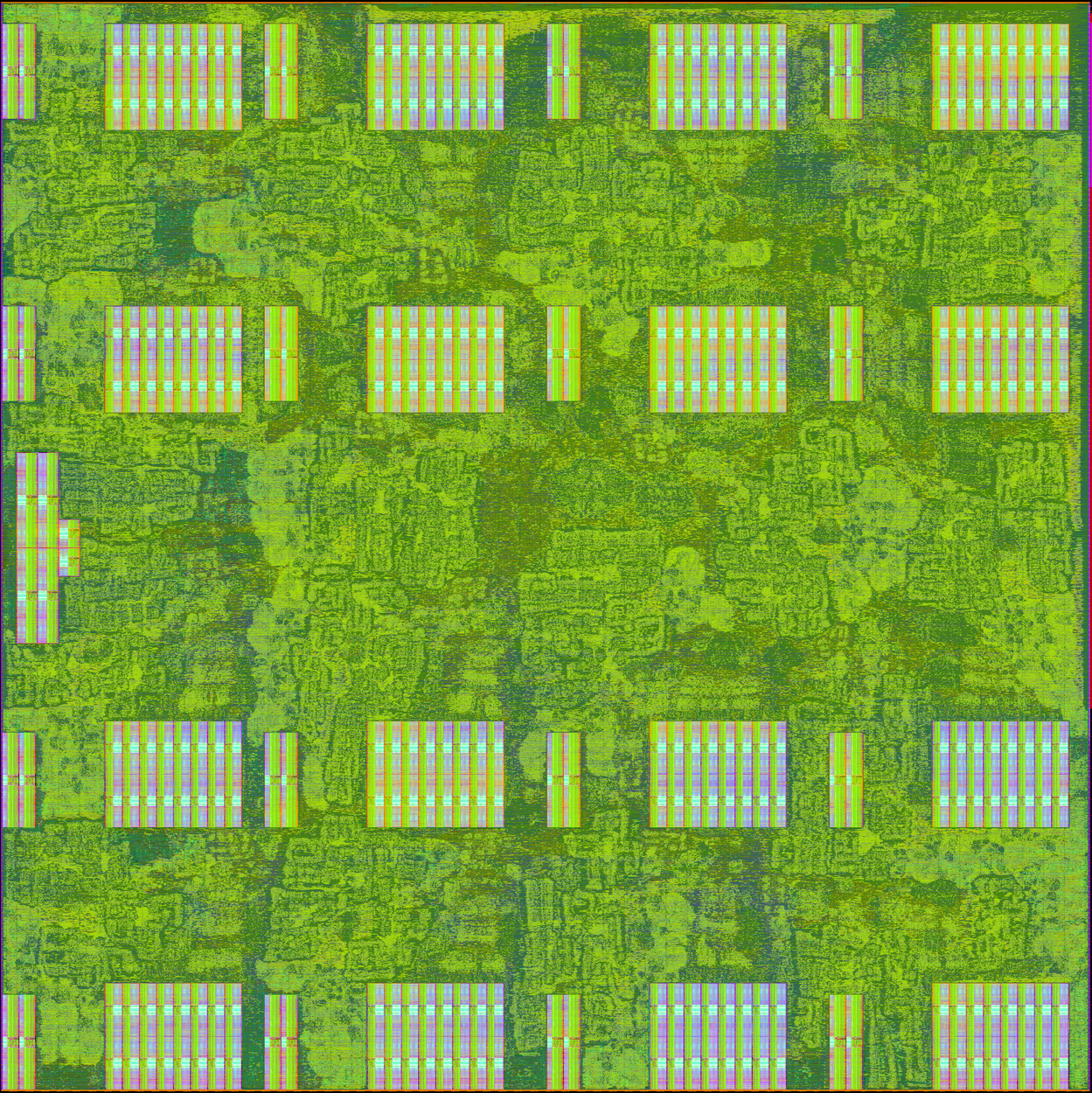}
    \subcaption{\mempoolspatz{64}{4}.}
    \label{fig:group_pnred_spatz}
  \end{minipage}\hfill%
  \begin{minipage}[t]{0.5\linewidth}
    \centering
    \includegraphics[width=0.89\linewidth]{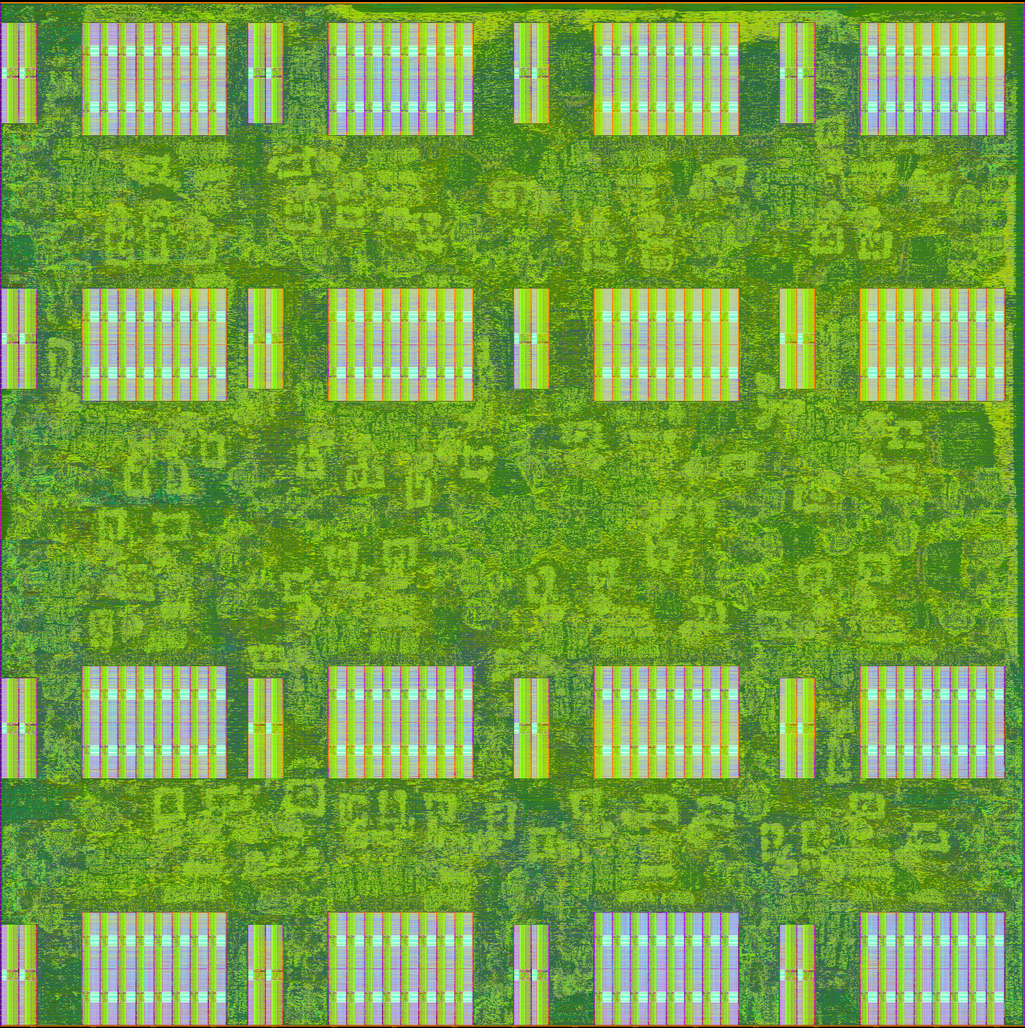}
    \subcaption{\mempool{256}.}
    \label{fig:group_pnred_snitch}
  \end{minipage}
  \caption{Placed-and-routed MemPool group instances of
    \mempoolspatz{64}{4} and \mempool{256}. Images to scale.}
  \label{fig:group_pnred}
\end{figure}

The \mempoolspatz{64}{4} group is \SI{27}{\percent} larger than the
\mempool{256} group. As discussed in \Cref{sec:phys-impl-1}, Spatz'
extra footprint is due to its \gls{VRF}. \Cref{fig:group_pnred} shows
the placed-and-routed \mempoolspatz{64}{4} and \mempool{256}
groups. The \mempoolspatz{64}{4} group of \Cref{fig:group_pnred_spatz}
was implemented as a \SI{1.87 x 1.87}{\milli\meter} macro, and the
\mempool{256} group of \Cref{fig:group_pnred_snitch} is a \SI{1.66 x
  1.66}{\milli\meter} macro.

The larger footprint has a small impact on MemPool's maximum operating
frequency, which drops from \SI{485}{\mega\hertz} for \mempool{256} to
\SI{471}{\mega\hertz} for \mempoolspatz{2}{128}. In typical
conditions, the three analyzed groups achieve the same operating
frequency of around \SI{590}{\mega\hertz}. The critical path of
\mempoolspatz{4}{64} is \num{53} gates long, going from a register at
the \gls{VLSU} boundary, through the \gls{VRF}'s read interface, and
through the \gls{VAU}, until reaching a register at the \gls{VRF}'s
write port. Overall, Spatz reaches similar frequencies to the rest of
the MemPool group despite the length of its critical path and does not
limit MemPool's frequency. Moreover, it is possible to pipeline Spatz'
critical path by adding a pipeline stage at the \gls{VRF} read
interface. The \mempoolspatz{128}{2} and \mempoolspatz{64}{4} groups
perform similarly in terms of footprint. Due to the reduced size of
the \spatz{4} \glspl{PE}, \mempoolspatz{64}{4}'s footprint is
approximately \SI{5}{\percent} smaller than \mempoolspatz{128}{2}'s.

Thanks to the comparable operating frequencies and much-improved
\gls{MACU} utilization, the Spatz-based MemPool systems achieve a peak
performance much higher than the Snitch-based MemPool. In particular,
\mempoolspatz{64}{4} reaches the highest performance,
\SI{285}{\giga\ops}, when running a \emph{matmul}$_{256}$,
\SI{71}{\percent} higher than \mempool{256}'s performance,
\SI{167}{\giga\ops}. Even considering the larger footprint of the
Spatz-based groups, \mempoolspatz{64}{4} reaches an area efficiency of
\SI{14.2}{\giga\ops\per\milli\meter\squared}, which is
\SI{33}{\percent} higher than \mempool{256}'s
\SI{10.6}{\giga\ops\per\milli\meter\squared}. Concerning the
Spatz-based MemPool instances, \mempoolspatz{128}{2} design reaches a
slightly lower area efficiency,
\SI{12.0}{\giga\ops\per\milli\meter\squared}, due to its lower
\emph{matmul}$_{256}$ performance and larger footprint compared to
\mempoolspatz{64}{4}.

\subsection{Power and Energy Efficiency}
\label{sec:energy-efficiency}

We used Synopsys' PrimePower 2022.03 to extract post-place-and-route
power results of the MemPool group in typical operating conditions
(TT/\SI{0.80}{\volt}/\SI{25}{\celsius}), using switching activities
extracted from a gate-level simulation of
\emph{matmul}$_{256}$. \Cref{tab:mempool_group} summarizes the power
results for the considered MemPool instances.

\begin{figure}[h]
  \centering
  \resizebox{\linewidth}{!}{
  \begin{tikzpicture}
    \begin{axis}[
      ybar stacked,
      ymin=0,
      ymax=1.5,
      bar width = 2em,
      x axis line style = {draw=none},
      tickwidth = 0pt,
      minor y tick num = 1,
      ylabel = {Power consumption [\si{\watt}]},
      ymajorgrids,
      yminorgrids,
      height = 7cm,
      nodes near coords,
      legend style={at={(0.5,-0.15)}, anchor=north, cells={anchor=west}},
      legend columns = 3,
      legend style={/tikz/every even column/.append style={column sep=0.1cm}},
      legend image code/.code={\draw [#1] (0cm,-0.1cm) rectangle (0.3cm,0.15cm);},
      symbolic x coords = {\mempool{256}, \mempoolspatz{128}{2}, \mempoolspatz{64}{4}},
      xtick distance = 1,
      every pin/.style={font=\footnotesize},
      every node near coord/.append style={xshift=8pt,anchor=west,font=\footnotesize},
      point meta = explicit symbolic]

      \addplot+[ybar, color=black, fill=color4!60, postaction={pattern=north east lines}] plot coordinates {(\mempool{256}, 0.177) [\SI{177}{\milli\watt}] (\mempoolspatz{128}{2}, 0.092) [\SI{92}{\milli\watt}] (\mempoolspatz{64}{4}, 0.044) [\SI{44}{\milli\watt}]};
      \addplot+[ybar, color=black, fill=color1!40, postaction={pattern=crosshatch dots}] plot coordinates {(\mempool{256}, 0.198) [\SI{198}{\milli\watt}] (\mempoolspatz{128}{2}, 0.343) [\SI{343}{\milli\watt}] (\mempoolspatz{64}{4}, 0.367) [\SI{367}{\milli\watt}]};
      \addplot+[ybar, color=black, fill=color2!60, postaction={pattern=north west lines}] plot coordinates {(\mempool{256}, 0.127) [\SI{127}{\milli\watt}] (\mempoolspatz{128}{2}, 0.113) [\SI{113}{\milli\watt}] (\mempoolspatz{64}{4}, 0.106) [\SI{106}{\milli\watt}]};
      \addplot+[ybar, color=black, fill=color3!60, postaction={pattern=dots}] plot coordinates {(\mempool{256}, 0.081) [\SI{81}{\milli\watt}] (\mempoolspatz{128}{2}, 0.087) [\SI{87}{\milli\watt}] (\mempoolspatz{64}{4}, 0.056) [\SI{56}{\milli\watt}]};
      \addplot+[ybar, color=black, fill=color5!60, postaction={pattern=crosshatch}] plot coordinates {(\mempool{256}, 0.514) [\SI{514}{\milli\watt}] (\mempoolspatz{128}{2}, 0.423) [\SI{423}{\milli\watt}] (\mempoolspatz{64}{4}, 0.338) [\SI{338}{\milli\watt}]};
      \addplot+[ybar, color=black, fill=color6!50] plot coordinates {(\mempool{256}, 0.197) [\SI{197}{\milli\watt}] (\mempoolspatz{128}{2}, 0.159) [\SI{159}{\milli\watt}] (\mempoolspatz{64}{4}, 0.159) [\SI{159}{\milli\watt}]};

      \legend{Snitch (without Register File), Register Files, \gls{MACU}, \acrlong{IDol}, Interconnect, L1 \gls{SPM}}
    \end{axis}
  \end{tikzpicture}}
  \caption{Power consumption breakdown of the \mempool{256},
    \mempoolspatz{128}{2}, and \mempoolspatz{64}{4} groups running the
    \emph{matmul}$_{256}$ kernel on typical conditions.}
  \label{fig:power_breakdown}
\end{figure}
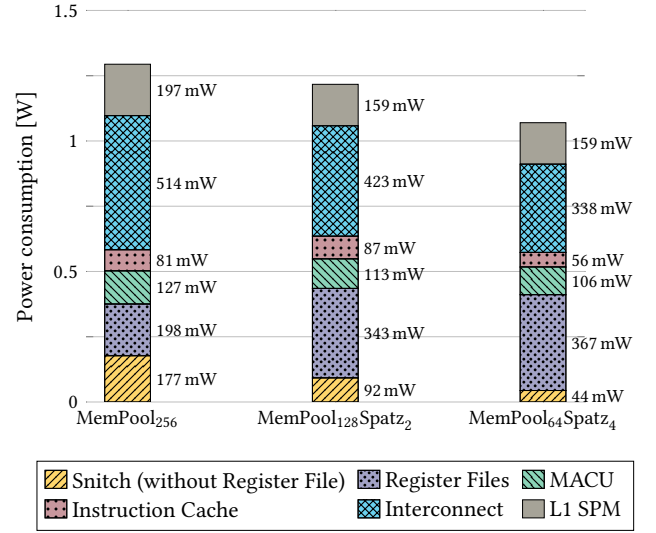

\mempool{256} consumes \SI{1.30}{\watt}, which is higher than the
consumption of both \mempoolspatz{128}{2} and
\mempoolspatz{64}{4}. \Cref{fig:power_breakdown} shows a power
breakdown of the three considered instances.

Snitch is a major contributor to the power consumption of
\mempool{256}, consuming \SI{375}{\milli\watt}, \SI{29}{\percent} of
the system's overall power consumption. More than half
(\SI{198}{\milli\watt}, \SI{53}{\percent}) of the power is consumed by
the register file alone. On the other hand, \mempool{256}'s
\glspl{MACU} consume only \SI{127}{\milli\watt}, comparable to the
power consumption of its instruction cache, \SI{81}{\milli\watt}. This
shows how the \gls{VNB} translates into a large energy overhead for
dispatching instruction. The unbalance in the power consumption
breakdown is even more striking because of the low power consumption
of the \glspl{MACU}, which consume only \SI{10}{\percent} of
\mempool{256}'s overall energy consumption. Finally, Snitch's data
reuse is limited by the size of its scalar register file. The largest
matrix blocks we could fit in the register file is \num{4x4}, \ie the
kernel loads eight elements from L1 for every $16$ multiply-accumulate
operations. Due to this high L1 memory traffic, the interconnects and
L1 \gls{SPM} \gls{SRAM} banks are responsible for most of
\mempool{256}'s power consumption, \SI{793}{\milli\watt}
(\SI{61}{\percent}).

Vector processing amortizes much of the power overheads spotted on
\mempool{256}. Snitch's power consumption is \SI{92}{\milli\watt} on
\mempoolspatz{128}{2} and \SI{44}{\milli\watt} on
\mempoolspatz{64}{4}, \SI{8}{\percent} and \SI{4}{\percent} of their
overall power consumption. Spatz' \gls{VRF} consumes a large portion
of the total power. On \mempoolspatz{128}{2}, the \gls{VRF} consumes
\SI{343}{\milli\watt}, \SI{30}{\percent} of the overall power
consumption. That number increases to \SI{367}{\milli\watt} on
\mempoolspatz{64}{4}, \SI{34}{\percent} of its total power
consumption. Although Spatz' \gls{VRF} consumes \SI{85}{\percent} more
power than Snitch's register files, its capacity is also four times
larger. Therefore, the largest matrix blocks we could fit in the
\gls{VRF} are \num{8x8}, \ie the kernel loads $16$ elements from L1
for every $64$ multiply-accumulate operations. As a result, the
\gls{VRF} acts as an L0 memory level, improving the locality and
reducing the rate at which the \glspl{PE} do expensive memory accesses
into the L1 \gls{SRAM} banks. In turn, this decreases the power
consumed by the interconnects and L1 \gls{SPM} banks, which consume
\SI{497}{\milli\watt} (\SI{46}{\percent}) on \mempoolspatz{64}{4}.

In terms of energy efficiency, Spatz is an highly viable \gls{PE}
option to build a shared-L1 cluster. It more than doubled the energy
efficiency of \mempool{256}, with \mempoolspatz{128}{2} reaching
\SI{234}{\giga\ops\per\watt} and \mempoolspatz{64}{4} reaching
\SI{266}{\giga\ops\per\watt}. Even considering Spatz' increased
footprint, for an area increase of \SI{27}{\percent} (mostly due to
the \gls{VRF}), compared to \mempool{256}, we increased the peak
performance by \SI{70}{\percent} and the energy efficiency by
\SI{107}{\percent}.


\section{Related Work}
\label{sec:related-work}

To the best of our knowledge, the tightly-coupled cluster of vector
processors architecture has not been explored in past literature.
Many vector processing units have been proposed in recent years,
thanks to new vector \glspl{ISA} such as Arm's
\gls{SVE}~\cite{Armv81M2019} and RISC-V's
\gls{RVV}~\cite{RISCV2022}. Examples of such large-scale vector
architectures based on the \gls{RVV} \gls{ISA} include BSC's
Vitruvius~\cite{Minervini2021}, PULP Platform's Ara~\cite{Ara2020},
and SiFive's P270~\cite{SiFiveP270} and X280~\cite{SiFiveX280}
cores. However, most of those units are large 64-bit vector processors
supporting double-precision floating-point operations, attached to
high-performance application-class scalar processors; hence they
achieve comparatively low efficiency due to the complex
micro-architecture of their supporting scalar cores~\cite{Zaruba2019}.

Small-scale vector units have been proposed for \glspl{FPGA}, where
the leanness of the vector processor is a constraint due to limited
\gls{FPGA} resources. Vicuna~\cite{Platzer2021} is a
timing-predictable vector processor compliant with \gls{RVV} version
0.10, synthesized on a Xilinx Series 7 \gls{FPGA}. Its \gls{VRF} was
implemented as a multi-ported \gls{RAM} due to concerns with timing
anomalies with a multi-banked \gls{VRF}. Vicuna's largest
configuration, comparable to \mempoolspatz{4}{4}, achieves up to
\SI{117}{\op\per\cycle} on an 8-bit \num{1024x1024} \emph{matmul}
kernel. Vicuna's multi-ported \gls{VRF} is similar to Spatz'
\gls{VRF}. To the best of our knowledge, there is no study about
Vicuna's scaling nor an \gls{ASIC} implementation of this
architecture, making a power or energy efficiency comparison with
Spatz difficult. The same can be said about other small-scale
\gls{RVV} vector units demonstrated on \glspl{FPGA}, \eg
Arrow~\cite{AlAssir2021} and AVA~\cite{Johns2020}.

Arm also started exploring embedded vector machines with its Helium
\gls{MVE}, an optional extension proposed as part of the Armv8.1-M
architecture~\cite{Armv81M2019}. The Arm
Cortex-M55~\cite{ArmCortexM552020} is the first processor to ship with
support to \gls{MVE}. However, to the best of our knowledge, no
quantitative assessment of the performance and efficiency of a
Cortex-M55 silicon implementation has been reported so far in the open
literature; hence a quantitative comparison with Spatz is not
possible. For what concerns a qualitative comparison, we observe that
the Helium \gls{MVE} defines eight $128$-bit wide vector registers as
aliases to the floating-point register file. On the M55 processor, the
$64$-bit datapath means Helium operates on a ``dual-beat regime,'' \ie
vector instructions execute in two cycles. This is enough to overlap
the execution of successive vector instructions in different
processing units without needing a superscalar core. However, the
scalar core needs to frequently issue instructions to the
Helium-capable processing unit to keep its pipeline busy. In contrast,
Spatz' longer vector registers and \gls{RVV}'s \gls{LMUL} register
grouping allow for a maximum vector length of $4096$ bits, keeping
\spatz{4} busy for $32$ cycles. This long execution, as shown by our
experimental results, massively amortizes the energy overhead of the
scalar core, which is a considerable part of the overall energy
consumption even running extremely data-parallel kernels such as the
matrix multiplication.


\section{Conclusion}
\label{sec:conclusion}

In this paper, we explored for the first time vector processing as an
option to build small and efficient \glspl{PE} for large-scale
clusters with tightly-coupled shared-L1 memory. We proposed Spatz, a
compact vector processing unit based on the \verb#Zve32x# embedded
integer subset of RISC-V's \gls{RVV} extension version 1.0. Spatz is
designed as a co-processor compliant with CORE-V's X-Interface generic
accelerator interface. In our case, Spatz was coupled with the Snitch
core. The most efficient Spatz configuration, \spatz{4}, has four
\glspl{MACU}, and a centralized latch-based \gls{VRF} with
\SI{2}{\kibi\byte}.

We implemented \spatz{4} with GlobalFoundries' 22FDX \gls{FDSOI}
technology and measured its energy consumption when running simple
arithmetic and memory operations. Spatz amortizes much of Snitch's
energy consumption on instruction fetching and decoding thanks to the
vector processing induced instruction fetch reduction, \ie by
mitigating the \gls{VNB}. While a small \spatz{4}-based cluster needs
\SI{7.9}{\pico\joule} to execute a multiply-accumulate elementary
operation, a Snitch-based cluster requires \SI{13.1}{\pico\joule},
\SI{66}{\percent} more energy for the same elementary operation, the
difference mainly due to the scalar core.

We also explored Spatz' \gls{PPA} when it is used as a \gls{PE} on
MemPool, a large-scale shared-L1 cluster with $256$ \glspl{MACU} and
\SI{1}{\mebi\byte} of L1. On a multi-core \spatz{4} environment,
\mempoolspatz{64}{4} reaches up to \SI{480}{\op\per\cycle}, a
\gls{MACU} utilization of \SI{94}{\percent}. This performance is much
higher than the performance achieved by the Snitch-based \mempool{256}
cluster, \SI{284}{\op\per\cycle}, \ie a \gls{MACU} utilization of
\SI{55}{\percent}. On the same kernel, \mempoolspatz{64}{4} consumes
\SI{1.07}{\watt}, \SI{18}{\percent} less than the \SI{1.30}{\watt}
consumed by \mempool{256} running the same kernel and operating
conditions. Spatz amortizes Snitch's power requirements, responsible
for \SI{29}{\percent} of \mempool{256}'s consumption. Moreover, its
\gls{VRF} improves locality and reduces accesses to the L1 \gls{SPM}.

In terms of energy efficiency, the \mempoolspatz{64}{4} instance
reaches \SI{266}{\giga\ops\per\watt}, more than twice the energy
efficiency reached by \mempool{256},
\SI{128}{\giga\ops\per\watt}. Even considering Spatz' impact on the
design's footprint, for an area increase of \SI{27}{\percent} (mostly
attributed to the \gls{VRF}), compared to \mempool{256}, we increased
the peak performance by \SI{70}{\percent} and the energy efficiency by
\SI{116}{\percent}. Spatz' agile vector architecture allows a highly
efficient \gls{PE}, which in turn improves the energy efficiency of a
vast array of architectures and ensures that it is computation and not
instruction fetch and decode to consume most of the power.

\bibliographystyle{ACM-Reference-Format}
\bibliography{spatz}

\end{document}